\documentclass[10pt,conference,english]{IEEEtran}
\usepackage[centertags]{amsmath}
\usepackage{amsfonts}
\usepackage{amssymb}
\usepackage{epsfig}
\usepackage{amsmath,  amssymb}
\usepackage{graphicx}
\usepackage[centertags]{amsmath}
\usepackage{clrscode}
\usepackage{cite}
\usepackage{fullpage}
\usepackage{amsmath,amssymb,mathrsfs,pseudocode}
\usepackage{color}
\usepackage[normalem]{ulem}
\usepackage{psfrag}
\usepackage{url}
\usepackage{babel}
\usepackage{authblk}
\usepackage{mathtools}
\usepackage[linesnumbered,ruled]{algorithm2e}
\usepackage[left=54pt, right=54pt, top=51pt, bottom=58pt]{geometry}
\usepackage{xspace}
\usepackage{multicol}
\usepackage{booktabs} 
\usepackage{multirow}
\usepackage[x11names,dvipsnames,table]{xcolor}

\usepackage{subfigure}

\usepackage{stackrel}
\usepackage{extarrows}

\newtheorem{theorem}{Theorem}
\newtheorem{lemma}{Lemma}

\newtheorem{definition}{Definition}
\newtheorem{remark}{Remark}

\usepackage{xspace}
\usepackage{bbm}
\input{mathlig}

\newcommand{\muspace}{\mspace{1mu}}

\DeclareRobustCommand{\scond}{\mathchoice{\muspace\vert\muspace}{\vert}{\vert}{\vert}}
\mathlig{|}{\scond}

\DeclareRobustCommand{\discint}{\mathchoice{\mspace{-1.5mu}:\mspace{-1.5mu}}{\mspace{-1.5mu}:\mspace{-1.5mu}}{:}{:}}
\mathlig{::}{\discint}

\newcommand{\Lc}{\mathcal{L}}
\newcommand{\Mc}{\mathcal{M}}

\newcommand{\Xc}{\mathcal{X}}
\newcommand{\Yc}{\mathcal{Y}}
\newcommand{\Zc}{\mathcal{Z}}

\newcommand{\Gcal}{\mathcal{G}}

\newcommand{\Pcal}{\mathcal{P}}

\newcommand{\Av}{{\bf A}}

\newcommand{\Xv}{{X}}

\newcommand{\Wv}{{\bf W}}

\newcommand{\dv}{{\bf d}}

\newcommand{\xv}{{x}}

\newcommand{\Vh}{{\hat{V}}}

\newcommand{\Zh}{{\hat{Z}}}

\newcommand{\dt}{{\tilde{d}}}

\def\textiid{i.i.d.\@\xspace}
\newcommand\iid{\ifmmode\text{ i.i.d. } \else \textiid \fi}

\def\clap#1{\hbox to 0pt{\hss#1\hss}}
\def\mathclap{\mathpalette\mathclapinternal}
\def\mathclapinternal#1#2{%
  \clap{$\mathsurround=0pt#1{#2}$}}

\let\oldstackrel\stackrel
\renewcommand{\stackrel}[2]{\oldstackrel{\mathclap{#1}}{#2}}

\newcommand{\indep}{\perp \!\!\! \perp}

\newcommand{\parastoo}{\textcolor{black}}

\newcommand{\Ni}{\textcolor{black}}

\newcommand{\Roy}{\textcolor{black}}

\newcommand{\Roycr}{\textcolor{black}}

\allowdisplaybreaks
\begin{document}
\title{Privacy-Utility Tradeoff in a Guessing Framework Inspired by Index Coding}

\author{Yucheng Liu$^{\dag}$, Ni Ding$^{*}$, Parastoo Sadeghi$^{\dag}$, and Thierry Rakotoarivelo$^{*}$\\\vspace{-0mm}
$^{\dag}$Research School of Electrical, Energy and Materials Engineering, Australian National University, Australia\\
$^{*}$Data61, Commonwealth Scientific and Industrial Research Organisation, Australia\\
Emails:  $^{\dag}$\{yucheng.liu, parastoo.sadeghi\}@anu.edu.au, $^{*}$\{ni.ding, thierry.rakotoarivelo\}@data61.csiro.au}


\maketitle

\begin{abstract}
This paper studies the tradeoff in privacy and utility in a single-trial multi-terminal guessing (estimation) framework using a system model that is inspired by index coding. There are $n$ independent discrete sources at a data curator. There are $m$ legitimate users and one adversary, each with some side information about the sources. The data curator broadcasts a distorted function of sources to legitimate users, which is also overheard by the adversary. In terms of utility, each legitimate user wishes to perfectly \parastoo{reconstruct} some of the unknown sources and
attain a certain gain in the estimation correctness for the remaining unknown sources.
In terms of privacy, the data curator wishes to minimize the \emph{maximal leakage}: the worst-case guessing gain of the adversary in estimating any target function of its unknown sources after receiving the broadcast data. Given the system settings, we derive fundamental performance lower bounds on the maximal leakage to the adversary, which are inspired by the notion of confusion graph and performance bounds for the index coding problem. We also detail a greedy privacy enhancing mechanism, which is inspired by the agglomerative clustering algorithms in the information bottleneck and privacy funnel problems.
\end{abstract}

\section{Introduction}\label{sec:intro}
\vspace{-0.125mm}

In this paper, we consider an information-theoretic multi-terminal guessing framework with side information. Our model is inspired by that of the index coding problem \cite{Birk--Kol2006,bar2011index}, but with a significant twist to place emphasis on privacy. 
Index coding is a communication problem where a sender aims to efficiently broadcast to multiple users through a noiseless channel. See \cite{arbabjolfaei2018fundamentals} and references therein. 
\Roycr{In particular, the security and privacy aspects of index coding have been investigated in \cite{dau2012security,ong2016secure,mojahedian2017perfectly,liu:vellambi:kim:sadeghi:itw18,narayanan2018private,liu2020secure,karmoose2019privacy}.}
Instead of \Roycr{maximizing the communication rate}, in our framework the sender's goal is to balance the privacy and utility performance in the broadcast, both measured based on the success rate of correctly estimating a certain parameter of interest about the sources \parastoo{through} a single guess. 
We consider multiple independent sources available at a data curator and assume that there are multiple legitimate users, as well as one adversary in the system. Each party (a user or the adversary) knows some sources a priori as side information. The data curator broadcasts \Roy{(discloses)} a distorted function of the sources to the users, which is also overheard by the adversary. 
\Roycr{Such framework has applications in various real-world scenarios, such as field data broadcasting from a paddock aggregator in the presence of a malicious agent in precision agriculture. }

For the adversary, we adopt the maximal leakage introduced in \cite{issa2019operational} as the privacy metric. \parastoo{It measures} the worst-case information leakage in terms of the gain of the \parastoo{adversary in maximum a posteriori estimation} of any target function of the unknown sources \emph{after and before} observing the disclosed data. For legitimate users, we define our utility metric \parastoo{such that it also reflects the improvement in users' guessing  ability}. 
Quite often in practice different sources are of different levels of priority to a user. 
We capture this by dividing the unknown sources of each user into two subsets.
Some essential source are required to be perfectly reconstructed by the user. That is, the correct guessing probability of such sources after observing the disclosed data is non-negotiable and must be $1$.
The remaining sources are less critical, for which the user requires the guessing gain \parastoo{about each such source} to be larger than a certain negotiable threshold. 
The privacy-utility tradeoff is cast as a constrained optimization problem, where the objective is to minimize the privacy leakage to the adversary, conditioned that \parastoo{the requirements on the utility} of the \parastoo{unknown sources for} the legitimate users are satisfied. 

The contributions of this paper are twofold. First, we derive two lower bounds on the privacy leakage given the source distribution and utility constraints (cf. Theorems \ref{thm:lower:bound:confusion}-\ref{thm:lower:bound:pm}) from different persepctives. 
These lower bounds serve as converse results (fundamental performance limits) for the privacy leakage.
\Roy{Second, we propose a greedy algorithm as \parastoo{the privacy mechanism} (cf. Algorithm \ref{alg:agglomerative}) inspired by the agglomerative clustering method used in the information bottleneck and privacy funnel problems \cite{slonim2000agglomerative,makhdoumi2014information,ding2019submodularity}.} 
We leverage the connection between our data disclosure problem and the index coding problem when investigating both the converse results and the greedy algorithm.



{\it Notation:} For non-negative integers $a$ and $b$, $[a]$ denotes the set $\{1,2,\cdots,a\}$ and $[a:b]$ denotes the set $\{a,a+1,\cdots,b\}$. 
For a set $S$, $|S|$ denotes its cardinality.
For any discrete random variable $Z$ with probability distribution $P_Z$, we denote its alphabet by $\Zc$ with realizations $z\in \Zc$. We denote an estimator of $Z$ by $\Zh$, whose alphabet is also $\Zc$.

\section{System Model and Problem Formulation}\label{sec:model}

Assume that a data curator observes $n$ independent discrete random sources $X_1,X_2,\ldots,X_n$. We assume a general distribution $P_{X_i}$ for each source. 
\Roy{Without loss of generality, we assume every source has full support.}
For brevity, when we say source $i$, we mean source $X_i$. For any $S\subseteq [n]$, set $S^c=[n]\setminus S$, $\Xv_S=(X_i: i\in S)$, $\xv_S=(x_i: i\in S)$, and $\Xc_S=\prod_{i\in S}\Xc_i$.
The data curator broadcasts a distorted version of $\Xv_{[n]}$, denoted by $Y$, generated according to the privacy mechanism $P_{Y|\Xv_{[n]}}$, to a number of legitimate users, which is also overheard by an adversary.

We consider a guessing framework where both the legitimate users and the adversary attempt to estimate a certain parameter $V$ of interest about the sources in a single trial.


Both the privacy and utility are measured in terms of a guessing gain $r$ as defined below.
Consider any party (a user or the adversary) that wishes to guess $V$ with the side information $Z$. 
Note that $V\indep Z$ due to source independence.
The party aims to maximize the correct guessing probability of $V$ upon observing $Y$ (i.e., the party employs the maximum \emph{a posteriori} estimator).
For each $(y,z)\in \Yc\times \Zc$,
we define the ratio between such maximized guessing probability after and before observing a $y \in \Yc$ given $z \in \Zc$ as
\begin{align}
r(V\to y|z)&\doteq  \frac{\max\limits_{P_{\Vh|Y=y,Z=z}} {\mathbb E}\big[ P_{\Vh|Y=y,Z=z}(V|y,z) \big]}{\max\limits_{P_{\Vh|Z=z}}{\mathbb E}\big[ P_{\Vh|Z=z}(V|z) \big]}   \label{eq:model:ratio}  \\
&=\frac{\max\limits_{v}P_{V|Y,Z}(\parastoo{v}|y,z)}{\max\limits_{v}P_{V}(v)}.  \label{eq:model:ratio:dddd}
\end{align}


%
%


{\bf \Roycr{Privacy Metric}:} We assume the adversary has side information $\Xv_{P}$ for some $P\subseteq [n]$, and is interested in a (possibly randomized) discrete function $U$ of the sources $\Xv_{Q}$ it does not know, where $Q\doteq P^c$. 
\Roy{Quite often in practice,} this function is chosen by the adversary and is unknown to the data curator. Therefore, we consider a worst-case privacy leakage measure, a conditional version of maximal leakage (MaxL) from \cite{issa2019operational}, as our privacy metric. 
\begin{definition}[Maximal Leakage,\cite{issa2019operational}]    \label{def:ML}
Given a finite discrete joint distribution $P_{\Xv_{[n]},Y}$, the maximal leakage from $\Xv_Q$ to Y given $\Xv_P$ is defined as
\begin{align}
&\Lc_{\rm max}(\hspace{-0.0625mm} \Xv_Q \hspace{-0.375mm} \to \hspace{-0.375mm} Y|\Xv_P \hspace{-0.0625mm})\doteq \hspace{-1mm} \sup_{U:U-\Xv_Q-(Y,\Xv_P)} \hspace{-1mm} \Lc(\hspace{-0.0625mm} U \hspace{-0.375mm}\to \hspace{-0.375mm} Y|\Xv_P \hspace{-0.0625mm}),   \label{eq:model:def:ml}
\end{align}
where
\begin{align}
\Lc(U\to Y|\Xv_P)\doteq \log {\mathbb E}_{P_{Y,\Xv_P}}\big[ r(U\to Y|\Xv_P) \big].  \label{eq:model:def:l}
\end{align}
\end{definition}

\begin{remark}
Note that the MaxL in Definition \ref{def:ML} assumes a different Markov chain model from \cite[Section III-E]{issa2019operational}: for our problem, the Markov chain model studied in \cite{issa2019operational} always \parastoo{reduces} to $U-\Xv_{[n]}-Y$ regardless of $Q$.
\end{remark}

For the rest of the paper, we refer to $\Lc_{\rm max}(\Xv_Q \to Y|\Xv_P)$ as $\Lc_{\rm max}$ when there is no ambiguity.
A computable expression of the MaxL in Definition \ref{def:ML} is presented as
\begin{align}
\Lc_{\rm max}=\log \sum_{y,\xv_P}\max_{\xv_{Q}}P_{Y,\Xv_P|\Xv_Q}(y,\xv_P|\xv_Q),   \label{eq:model:ml:sibson}
\end{align}
which can be obtained following a similar approach to \cite{issa2019operational}.
We omit the derivation details due to limited space.

{\bf \Roycr{Utility Constraints}:} There are $m$ legitimate users. User $i\in [m]$ knows some sources $\Xv_{A_i}$ a priori as side information for some $A_i\subseteq [n]$, and is interested in all the remaining sources $\Xv_{A_i^c}$, 
which are divided into two groups of different levels of priority:
\begin{itemize}
\item Source $\Xv_{W_i}$ for some $W_i\subseteq A_i^c$ are indispensable to the user, and thus must be correctly guessed by user $i$ with probability of $1$ (i.e., perfect decoding).
\item The rest of the sources, $\Xv_{G_i}$, where $G_i\doteq A_i^c\setminus W_i$, are less essential yet still useful/interesting. Thus, user $i$ requires the guessing ability gain upon observing $Y$ to be larger than a certain threshold $d_i$.
\end{itemize}
These result in the following two kinds of utility constraints. For any $i\in [m]$, we have
\begin{alignat}{2}
&H(\Xv_{W_i}|Y,\Xv_{A_i})=0,                     && \qquad  \forall i\in [m],   \label{eq:model:utility:wi}   \\
&D(\Xv_{G_i}\to Y|\Xv_{A_i})\ge d_i,          && \qquad  \forall i\in [m],   \label{eq:model:utility:gi}
\end{alignat}
where $D(\Xv_{G_i}\to Y|\Xv_{A_i})$ is defined as
\begin{align}
&D(\Xv_{G_i}\to Y|\Xv_{A_i}) \hspace{-0.25mm} \doteq \hspace{-0.25mm} {\mathbb E}_{P_{Y,\Xv_{A_i}}} \hspace{-0.5mm} \big[ \log r(\Xv_{G_i} \hspace{-0.5mm} \to \hspace{-0.5mm} Y|\Xv_{A_i}) \big].      \label{eq:model:utility:leakage:def}
\end{align}

{Note that for constraint \eqref{eq:model:utility:gi}, each legitimate user $i$ is interested in obtaining the source $\Xv_{G_i}$ \Roy{rather than} a function/feature of $\Xv_{G_i}$. 
%
We simplify the notation $D(\Xv_{G_i}\to Y|\Xv_{A_i})$ to $D_i$ when there is no ambiguity.
}


\begin{remark}
Note the subtle difference between $D_i$ and $\Lc(\Xv_{G_i}\to Y|\Xv_{A_i})=\log {\mathbb E}_{P_{Y,\Xv_{A_i}}}\big[ r(\Xv_{G_i}\to Y|\Xv_{A_i}) \big]$ as in \eqref{eq:model:def:l}. 
The latter is lower bounded by the former due to Jensen's inequality.
From the data curator's viewpoint, requesting $D_i$ to be above a certain threshold is more stringent than requesting $\Lc(\Xv_{G_i}\to Y|\Xv_{A_i})$ to be above that threshold.
We use $D_i$ rather than $\Lc(\Xv_{G_i}\to Y|\Xv_{A_i})$ as our utility measure as it leads to a simple closed-form result characterizing $\Lc_{\rm max}$ in terms of $D_i$ (cf. Lemma \ref{lem:lower:bound:core}). 
\end{remark}

\begin{remark}  \label{rem:model:utility}
The two types of utility constraints in \eqref{eq:model:utility:wi} and \eqref{eq:model:utility:gi} can be unified to be represented in terms of the same function $D$: The perfect decoding constraint \eqref{eq:model:utility:wi} is equivalent to requiring that 
\begin{align}
D(\Xv_{W_i}\to Y|\Xv_{A_i})&={\mathbb E}_{P_{Y,\Xv_{A_i}}}  \big[ \log \frac{1}{\max\limits_{\xv_{W_i}}P_{\Xv_{W_i}}(\xv_{W_i})} \big]  \nonumber   \\
&=H_{\infty}(\Xv_{W_i}),   \label{eq:model:utility:remark}
\end{align}
where $H_{\infty}(\Xv_{W_i})$ denotes the min-entropy (i.e., R\'enyi entropy of order $\infty$ \cite{renyi1961measures}) of $\Xv_{W_i}$. \parastoo{One} can show that for any $i\in [m]$, $D_i\le H_{\infty}(\Xv_{G_i})$. Consequently, 
we always require that $0\le d_i\le H_{\infty}(\Xv_{G_i})$ for any $i\in [m]$.
\end{remark}

{\bf Privacy-Utility Tradeoff:}
We denote such system by the 5-tuple
$(P_{\Xv_{[n]}},\Av,\Wv,\dv,P)$, where $\Av \doteq (A_i,i\in [m])$, $\Wv \doteq (W_i,i\in [m])$, and $\dv \doteq (d_i,i\in [m])$. Note that $G_i$ is determined by $W_i $ and $A_i$, and $Q$ is determined by $P$.

To design the privacy mechanism $P_{Y|\Xv_{[n]}}$, we need to consider the fundamental tradeoff between the privacy and utility. 
Any data distortion that reduces the information leakage to the adversary can decrease the utility obtained by the users.
Such tradeoff is formulated by the following constrained optimization problem.
\begin{align}
\inf_{\substack{P_{Y|\Xv_{[n]}}\in \\ \Pcal_{Y|\Xv_{[n]}}(P_{\Xv_{[n]}},\Av,\Wv,\dv,P)}}\Lc_{\rm max}(\Xv_Q\to Y|\Xv_P),    \label{eq:model:tradeoff}
\end{align}
where $\Pcal_{Y|\Xv_{[n]}}(P_{\Xv_{[n]}},\Av,\Wv,\dv,P)$ denotes the collection of randomized mappings $P_{Y|\Xv_{[n]}}$ that satisfy \eqref{eq:model:utility:wi} and \eqref{eq:model:utility:gi} for the problem $(P_{\Xv_{[n]}},\Av,\Wv,\dv,P)$.


Due to the non-convexity of \eqref{eq:model:tradeoff}, instead of providing an explicit solution, we derive lower bounds on $\Lc_{\rm max}$ by taking inspiration from index coding. 
These bounds serve as fundamental performance limits that cannot be surpassed \parastoo{by any mechanism} because they are enforced by the system $(P_{\Xv_{[n]}},\Av,\Wv,\dv,P)$. 
\Roy{In Section \ref{sec:privacy:mechanism}, we design an achievable mechanism based on the idea of agglomerative clustering.}
\newcommand{\Ae}{P}
\newcommand{\Ge}{Q}

\section{Lower Bounds on the Privacy Leakage}  \label{sec:lower:bound}

We derive two information-theoretic lower bounds on the privacy leakage. One is based on the utility constraint \eqref{eq:model:utility:wi} only, while the other is obtained based on both \eqref{eq:model:utility:wi} and \eqref{eq:model:utility:gi}. 

\subsection{Lower Bound Based on the Confusion Graph}  \label{sec:lower:bound:confusion}

The utility constraint \eqref{eq:model:utility:wi} indicates that for user $i$, any different realizations $\xv_{W_i}\neq \xv_{W_i}'\in \Xc_{W_i}$ must be distinguishable based on the released $y\in \Yc$, as well as the user's side information $\xv_{A_i}\in \Xc_{A_i}$. 
To describe such distinguishability, 
we recall the notion of confusion graph for index coding \cite{alon2008broadcasting}.

\begin{definition}[Confusion graph\cite{alon2008broadcasting}]
Any two realizations of the $n$ sources $\xv_{[n]}^1,\xv_{[n]}^2 \in \Xc_{[n]}$ are \emph{confusable} if there exists some user $i\in [m]$ such that $\xv_{W_i}^1\neq \xv_{W_i}^2$ and $\xv_{A_i}^1=\xv_{A_i}^2$. A confusion graph $\Gamma$ is an undirected uncapacitated graph with $|\Xc_{[n]}|$ vertices such that every vertex corresponds to a unique realization $\xv_{[n]}\in \Xc_{[n]}$ and an edge connects two vertices if and only if their corresponding realizations are confusable.
\end{definition}

Therefore, to ensure \eqref{eq:model:utility:wi}, a group of realizations of $\Xv_{[n]}$ can be mapped to the same $y$ with nonzero probability only if they are pairwisely not confusable.
More rigorously, for any $S\subseteq [n]$, define
\begin{alignat}{2}
\Yc(\xv_S)&\doteq \{ y\in \Yc:P_{Y|\Xv_S}(y|\xv_S)>0 \}, &&\qquad \forall \xv_S\in \Xc_S,   \nonumber   \\
\Xc_S(y)&\doteq \{ \xv_S\in \Xc_S:P_{Y|\Xv_S}(y|\xv_S)>0 \}, &&\qquad \forall y\in \Yc.  \nonumber
\end{alignat}
Then, we have the following lemma. We omit the proof as it can be simply done by contradiction. 
\begin{lemma}    \label{lem:confusion}
Given a $P_{Y|\Xv_{[n]}}$ satisfying \eqref{eq:model:utility:wi}, for any two confusable $\xv_{[n]}^1,\xv_{[n]}^2 \in \Xc_{[n]}$, we have $\Yc(\xv_{[n]}^1)\cap \Yc(\xv_{[n]}^2)=\emptyset. $
\end{lemma}


Given a set $S\subseteq [n]$ and a specific realization $\xv_S\in \Xc_S$, we define $\Gamma(\xv_S)$ as the subgraph of $\Gamma$ induced by all the vertices $\xv_{[n]}$ such that $\xv_{[n]}=(\xv_S,\xv_{S^c})$ for some $\xv_{S^c}\in \Xc_{S^c}$. 
Notice that for any $\xv_S^1\neq \xv_S^2\in \Xc_S$ and $\xv_{S^c}^1\neq \xv_{S^c}^2\in \Xc_{S^c}$, $(\xv_S^1,\xv_{S^c}^1)$ and $(\xv_S^1,\xv_{S^c}^2)$ are confusable if and only if $(\xv_S^2,\xv_{S^c}^1)$ and $(\xv_S^2,\xv_{S^c}^2)$ are confusable. Hence, given $S\subseteq [n]$, the subgraphs $\Gamma(\xv_S)$, $\xv_S\in \Xc_S$ are isomorphic to each other, and thus we simply denote any such subgraph by $\Gamma(S)$. 



We present our main result of this subsection as follows. 
\begin{theorem}   \label{thm:lower:bound:confusion}
For the problem \eqref{eq:model:tradeoff} with confusion graph $\Gamma$:
\begin{align}
\Lc_{\rm max}\ge \log \omega(\Gamma(P)),   \label{eq:lower:bound:confusion}
\end{align}
where $\omega(\cdot)$ is the clique number (size of the largest clique) \parastoo{of a graph}. 
\end{theorem}
\begin{IEEEproof}
Consider any $\xv_P\in \Xc_P$. There exists some realizations $\xv_Q^1,\xv_Q^2,\ldots,\xv_Q^{\omega(\Gamma(\xv_P))} \in \Xc_Q$ whose corresponding vertices in the subgraph $\Gamma(\xv_P)$ form a clique, 
which indicates that the realizations $(\xv_P,\xv_Q^1),(\xv_P,\xv_Q^2),\ldots,(\xv_P,\xv_Q^{\omega(\Gamma(\xv_P))})$ also form a clique in $\Gamma$. 
That is, the realizations $(\xv_P,\xv_Q^1),\ldots,(\xv_P,\xv_Q^{\omega(\Gamma(\xv_P))})$ are pairwisely confusable. 
Then by Lemma \ref{lem:confusion}, for any $k\neq k'\in [\omega(\Gamma(\xv_P))]$, we have
\begin{align}
\Yc( (\xv_P,\xv_Q^k) )\cap \Yc( (\xv_P,\xv_Q^{k'}) )=\emptyset.    \label{eq:thm:confusion:disjoint}
\end{align}

%

Therefore, we have
\begin{align}
&\sum_{y} \max_{\xv_Q} P_{Y|\Xv_{[n]}}(y|\xv_P,\xv_Q)    \nonumber   \\
&\ge \sum_{k\in [\omega(\Gamma(\xv_P))]} \sum_{y\in \Yc(\xv_P,\xv_Q^k)} \max_{\substack{\xv_Q}} P_{Y|\Xv_{[n]}}(y|\xv_P,\xv_Q)   \nonumber   \\
&\ge \sum_{k\in [\omega(\Gamma(\xv_P))]} \sum_{y\in \Yc(\xv_P,\xv_Q^k)} P_{Y|\Xv_{[n]}}(y|\xv_P,\xv_Q^k)   \nonumber   \\
&=\sum_{k\in [\omega(\Gamma(\xv_P))]} 1=\omega(\Gamma(\xv_P))=\omega(\Gamma(P)),    \label{eq:thm:confusion:eachxp}
\end{align}
where the first inequality follows from \eqref{eq:thm:confusion:disjoint}. 
Hence, we have
\begin{align}
\Lc_{\rm max}
&=\log \sum_{\xv_P} P_{\Xv_P}(\xv_P) \sum_{y\in \Yc} \max_{\xv_Q} P_{Y|\Xv_{[n]}}(y|\xv_P,\xv_Q)      \nonumber  \\
&\ge \log \sum_{\xv_P} P_{\Xv_P}(\xv_P)\cdot \omega(\Gamma(P))=\log \omega(\Gamma(P)),   \nonumber  
\end{align}
where the first equality is due to source independence, 
and the inequality follows from \eqref{eq:thm:confusion:eachxp}. 
\end{IEEEproof}

\subsection{Lower Bound Based on Guessing Gain and \parastoo{Polymatroidal Functions}}

We introduce a key lemma that serves as the baseline in the lower bound to be developed in this subsection. 
\begin{lemma}  \label{lem:lower:bound:core}
For the problem \eqref{eq:model:tradeoff},
we have
\begin{align}
&\Lc_{\rm max}\ge \max\{ I(\Xv_Q;Y|\Xv_P), \max_{i\in [m]:G_i\subseteq Q}\Delta_i \},   \label{eq:lower:bound:core}
\end{align}
where $\Delta_i\doteq D_i+H(\Xv_{W_i\cap Q})+I(\Xv_{A_i\cap Q};Y|\Xv_{P}). $
\end{lemma}
\vspace{0.5mm}

The proof is presented in Appendix \ref{appendix:lem:core}.

For a given system $(P_{\Xv_{[n]}},\Av,\Wv,\dv,P)$, $H(\Xv_{W_i\cap Q})$ has a fixed value 
and $D_i$ is lower bounded by $d_i$ according to \eqref{eq:model:utility:gi}.
Hence, the only terms in \eqref{eq:lower:bound:core} that still depend on $P_{Y|\Xv_{[n]}}$ are the mutual information $I(\Xv_P;Y|\Xv_Q)$ and $I(\Xv_{A_i\cap Q};Y|\Xv_P)$. 
We further bound these mutual information terms below. 



We draw inspiration from the polymatroidal bound \cite{blasiak2011lexicographic,Arbabjolfaei--Bandemer--Kim--Sasoglu--Wang2013} for index coding. 
The bound is based on the polymatroidal axioms, which capture Shannon-type inequalities on the entropy function and play a central role in computing converse results in network information theory \cite{yeung2008information}.

\begin{lemma}\label{lem:lower:bound:pm}
Consider the system $(P_{\Xv_{[n]}},\Av,\Wv,\dv,P)$. For any disjoint $V,Z\subseteq [n]$, we have
\begin{align}
I(\Xv_V;Y|\Xv_Z)=g(Z^c)-g(Z^c\cap V^c),
\end{align}
for some polymatroidal rank function $g(S),S\subseteq [n]$ such that for any $i\in [n]$, $W\subseteq W_i$, $G\subseteq W^c\cap A_i^c$,
\begin{align}
&H(\Xv_W)=g(G\cup W)-g(G),     \label{eq:lower:bound:lem:pm:rate}
\end{align}
and
\begin{align}
&g(\emptyset)=0,      \label{eq:lower:bound:lem:pm:nonnegativity}  \\
&g(S')\ge g(S),    \qquad \text{if $S\subseteq S'$},          \label{eq:lower:bound:lem:pm:monotonicity}   \\
&g(S')+g(S)\ge g(S'\cup S)+g(S'\cap S).        \label{eq:lower:bound:lem:pm:submodularity}
\end{align}
\end{lemma}

\begin{IEEEproof}
Define $g(S)\doteq H(Y|\Xv_{S^c})-H(Y|\Xv_{[n]}),\forall S\subseteq [n]$. We have
$I(\Xv_V;Y|\Xv_Z)=g(Z^c)-g(Z^c\cap V^c).$ 
It remains to show that this $g(S)$ satisfies \eqref{eq:lower:bound:lem:pm:rate}-\eqref{eq:lower:bound:lem:pm:submodularity}.

For \eqref{eq:lower:bound:lem:pm:rate}, consider any $i\in [n]$, $W\subseteq W_i$, $G\subseteq W^c\cap A_i^c$. Set $A=[n]\setminus W\setminus G$, and one can verify that $A_i\subseteq A$. Hence,
\begin{align}
H(\Xv_W)&=H(\Xv_W|\Xv_A)-H(\Xv_W|Y,\Xv_A)    \label{eq:lower:bound:lem:pm:proof:1}  \\
&=H(Y|\Xv_A)-H(Y|\Xv_W,\Xv_A)    \nonumber   \\
&=g(W,G)-g(G),    \nonumber
\end{align}
where \eqref{eq:lower:bound:lem:pm:proof:1} is due to source independence, \eqref{eq:model:utility:wi}, and $A_i\subseteq A$.

For \eqref{eq:lower:bound:lem:pm:nonnegativity}, we have $g(\emptyset)=H(Y|\Xv_{[n]})-H(Y|\Xv_{[n]})=0$.

For \eqref{eq:lower:bound:lem:pm:monotonicity}, for any $S\subseteq S'\subseteq [n]$, $S'^c\subseteq S^c$, and thus
\begin{align}
g(S')=H(Y|\Xv_{S'^c})\ge H(Y|\Xv_{S^c})=g(S).   \nonumber
\end{align}

For \eqref{eq:lower:bound:lem:pm:submodularity}, consider any $S,S'\subseteq [n]$. Set $S^c\cap S'^c=S_0$, $S^c\setminus S_0=S_1$, and $S'^c\setminus S_0=S_2$. 
We have
\begin{align}
&g(S')+g(S)    \nonumber  \\
&=H(Y, \hspace{-0.375mm} \Xv_{S_0\cup S_2} \hspace{-0.25mm} ) \hspace{-0.375mm} + \hspace{-0.375mm} H(Y, \hspace{-0.375mm} \Xv_{S_0\cup S_1} \hspace{-0.25mm} ) \hspace{-0.375mm} - \hspace{-0.375mm} H( \hspace{-0.25mm} \Xv_{S_0\cup S_2} \hspace{-0.25mm} ) \hspace{-0.375mm} - \hspace{-0.375mm} H( \hspace{-0.25mm} \Xv_{S_0\cup S_1} \hspace{-0.25mm} )    \nonumber    \\
&\ge H(Y, \hspace{-0.25mm} \Xv_{S_0} \hspace{-0.375mm} ) \hspace{-0.375mm} + \hspace{-0.375mm} H(Y, \hspace{-0.25mm} \Xv_{S_0\cup S_1\cup S_2} \hspace{-0.375mm} ) \hspace{-0.375mm} - \hspace{-0.375mm} H(\Xv_{S_0} \hspace{-0.375mm} ) \hspace{-0.375mm} - \hspace{-0.375mm} H(\Xv_{S_0\cup S_1\cup S_2} \hspace{-0.375mm} )    \nonumber    \\
&=g(S\cup S')+g(S\cap S'),   \nonumber
\end{align}
where the inequality follows from the submodularity of the entropy function, as well as source independence.
\end{IEEEproof}

The above bound can be solved using either linear programming (LP) or Fourier-Motzkin elimination \cite[Appendix D]{elgamal_yhk}. 
In the system $(P_{\Xv_{[n]}},\Av,\Wv,\dv,P)$, for any disjoint $V,Z\subseteq [n]$, let
\begin{align*}
\Lambda(V,Z)\doteq \Roycr{\min_{\substack{\text{$g$ satisfying \eqref{eq:lower:bound:lem:pm:rate}-\eqref{eq:lower:bound:lem:pm:submodularity}}}}\{ g(Z^c)-g(Z^c\cap V^c) \}}.
\end{align*}
Combining Lemmas \ref{lem:lower:bound:core} and \ref{lem:lower:bound:pm} gives the following result.

\begin{theorem}\label{thm:lower:bound:pm}
For the problem \eqref{eq:model:tradeoff}, 
we have
\begin{align}
&\Lc_{\rm max}\ge \max\{ \Lambda(Q,P),  \nonumber  \\
&\max_{\substack{i\in [m]:G_i\subseteq Q}}\big( d_i+H(\Xv_{W_i\cap Q})+\Lambda(A_i\cap Q,P) \big) \}.   \nonumber 
\end{align}
\end{theorem}


\begin{remark}
\parastoo{In the index coding problem, we usually assume uniformly distributed independent sources and a deterministic mapping $P_{Y|\Xv_{[n]}}$. In contrast, Theorems \ref{thm:lower:bound:confusion} and \ref{thm:lower:bound:pm} generally hold for any discrete independent source distribution and make no assumption on the privacy mechanism.}
\end{remark}

In general, Theorems \ref{thm:lower:bound:confusion} and \ref{thm:lower:bound:pm} can outperform each other. 

\section{Privacy Mechanism Design}   \label{sec:privacy:mechanism}

We develop a greedy algorithm to provide a solution for the problem \eqref{eq:model:tradeoff}.
The algorithm is based on the agglomerative clustering method, which has been used in the information bottleneck \cite{slonim2000agglomerative} and the privacy funnel problem \cite{makhdoumi2014information}.

Consider a given system $(P_{\Xv_{[n]}},\Av,\Wv,\dv,P)$. To design the privacy mechanism $P_{Y|\Xv_{[n]}}$, we start from the one-to-one deterministic mapping with \parastoo{$\Yc = \mathcal \Xv_{[n]}$ and $P_{Y|\Xv_{[n]}}(y|\xv_{[n]}) = 1$ iff $y=\xv_{[n]}$}\footnote{Note that such one-to-one mapping allows every user to perfectly reconstruct every source and thus satisfies \eqref{eq:model:utility:wi} and \eqref{eq:model:utility:gi} for sure. Nevertheless, it also leads to the largest privacy leakage as the adversary can also perfectly reconstruct $\Xv_Q$ and subsequently any function $U$ it is interested in.} 
and then iteratively merge some elements of $\Yc$ to make the privacy leakage smaller (in other words, we ``blur" the revealed information), while still ensuring the utility for the users at an acceptable level, i.e., satisfying \eqref{eq:model:utility:wi} and \eqref{eq:model:utility:gi}. \Roy{In particular, to ensure \eqref{eq:model:utility:wi}, we again utilize the notion of confusion graph\cite{alon2008broadcasting} introduced in Section \ref{sec:lower:bound:confusion}.}

%
%
%

Based on the merging idea discussed above, we propose an agglomerative clustering algorithm in Algorithm \ref{alg:agglomerative}. 
Let $Y^{y_1,y_2}$ be the resulting random variable from merging any $y_1,y_2\in \Yc$.
{
Let $\Theta$ denote the collection of $\{y_1,y_2\}$ such that merging them does not violate \eqref{eq:model:utility:wi} and \eqref{eq:model:utility:gi} and strictly reduces the privacy leakage to the adversary: 
\begin{equation} \label{eq:Theta}
\small
    \begin{aligned}
        \Theta \doteq  & \big\{ \{y_1,y_2\} \in \Yc \times \Yc \colon y_1 \neq y_2, \text{ any two } \\
                 & \  \xv_{[n]},\xv_{[n]}'\in \Xc_{[n]}(y_1)\cup \Xc_{[n]}(y_2) \text{ are not confusable,} \\
                 & \ D(\Xv_{G_i}\to Y^{y_1,y_2}|\Xv_{A_i})\ge d_i, \forall i \in [m], \\
                 & \ \Lc_{\rm max}(\Xv_Q\to Y^{y_1,y_2}|\Xv_P)< \Lc_{\rm max}(\Xv_Q\to Y|\Xv_P)\big\}. 
    \end{aligned}
\end{equation}
The algorithm terminates if $\Theta$ becomes an empty set. 

        \begin{algorithm} \label{alg:agglomerative}
	       \small
	       \SetAlgoLined
	       \SetKwInOut{Input}{input}\SetKwInOut{Output}{output}
	       \SetKwFor{For}{for}{do}{endfor}
            \SetKwRepeat{Repeat}{repeat}{until}
            \SetKwIF{If}{ElseIf}{Else}{if}{then}{else if}{else}{endif}
	       \BlankLine
           \Input{The system $(P_{\Xv_{[n]}},\Wv,\Av,\dv,P)$.}
	       \Output{Privacy-preserving mechanism $P_{Y|\Xv_{[n]}}$.}
	       \BlankLine
            Initialization: $\Yc \leftarrow \Xc_{[n]}$, $P_{Y|\Xv_{[n]}}(y|\xv_{[n]}) \leftarrow 1$ iff $y=\xv_{[n]}$ and obtain $\Theta$ based on $\Yc$ by \eqref{eq:Theta}\label{step:initialization}\;
            \Repeat{$\Theta = \emptyset$}{
                $\{y_1^*,y_2^*\} \leftarrow \arg\min_{\{y_1,y_2\} \in \Theta} \mathcal{L}_{\rm max}(\Xv_Q\to Y^{y,y'}|\Xv_P)$\label{step:Min}\;
                Merge $y_1^*$ and $y_2^*$ into $\bar{y}$: $ \bar{y} \leftarrow \{y_1^*,y_2^*\}$\;
                Obtain the new $Y$ by letting $\Yc \leftarrow \Yc\setminus \{y_1^*,y_2^*\}\cup \{{\bar y}\}$ and $P_{Y|\Xv_{[n]}}({\bar y}|\xv_{[n]}) \leftarrow P_{Y|\Xv_{[n]}}(y_1^*|\xv_{[n]})+P_{Y|\Xv_{[n]}}(y_2^*|\xv_{[n]})$ for any $\xv_{[n]}\in \Xc_{[n]}$ while keeping the rest of $P_{Y|\Xv_{[n]}}$ unchanged\;
                Obtain the new $\Theta$ by \eqref{eq:Theta} based on updated \parastoo{$Y$}\label{step:Theta}\;
            }

            \Return $P_{Y|\Xv_{[n]}}$\;
	   \caption{Agglomerative clustering algorithm for problem \eqref{eq:model:tradeoff}}
	   \end{algorithm}

\begin{remark}
To compute Algorithm \ref{alg:agglomerative} more efficiently, notice that finding $\arg\min_{\{y_1,y_2\} \in \Theta}L_{\rm max}(\Xv_Q\to Y^{y_1,y_2}|\Xv_P)$ is equivalent to finding $\arg\max_{\{y_1,y_2\} \in \Theta}\big(2^{\Lc_{\rm max}(\Xv_Q\to Y|\Xv_P)}-2^{\Lc_{\rm max}(\Xv_Q\to Y^{y_1,y_2}|\Xv_P)\big)}$ in step~\ref{step:Min}, which can be computed as
\begin{align}
&2^{\Lc_{\rm max}(\Xv_Q\to Y|\Xv_P)}-2^{\Lc_{\rm max}(\Xv_Q\to Y^{y_1,y_2}|\Xv_P)}   \nonumber   \\
&=\sum_{\xv_P}\max_{\xv_Q}P_{\Xv_P}(\xv_P)\cdot P_{Y|\Xv_{[n]}}(y_1|\xv_P,\xv_Q)   \nonumber   \\
&\quad +\sum_{\xv_P}\max_{\xv_Q}P_{\Xv_P}(\xv_P)\cdot P_{Y|\Xv_{[n]}}(y_2|\xv_P,\xv_Q)   \nonumber   \\
&\quad -\sum_{\xv_P}\max_{\xv_Q}P_{\Xv_P}(\xv_P)\cdot P_{Y|\Xv_{[n]}}({\bar y}|\xv_P,\xv_Q)   \nonumber   \\
&=\sum_{\xv_P\in \Xc_P(y_1)}P_{\Xv_P}(\xv_P)\cdot 1+\sum_{\xv_P\in \Xc_P(y_2)}P_{\Xv_P}(\xv_P)\cdot 1   \nonumber   \\
&\quad -\sum_{\xv_P\in \Xc_P({\bar y})}P_{\Xv_P}(\xv_P)\cdot 1      \nonumber   \\
&=P_{\Xv_P}(\Xc_P(y_1))+P_{\Xv_P}(\Xc_P(y_2))-P_{\Xv_P}(\Xc_P({\bar y})).    \nonumber
\end{align}
\end{remark}

\section{Concluding Remarks}\label{sec:conclusion}


To evaluate the performance of our main results, we randomly generate the system $(P_{\Xv_{[n]}},\Av,\Wv,\dv,P)$ by $500$ times according to the following conditions: 
\begin{itemize}
\item $n=m=5$, $W_i=\{i\}$ for any user $i\in [5]$, and $\Av$ is generated based on a randomly chosen graph $\Gcal$ from the $9608$ nonisomorphic 5-vertex directed graphs \cite{arbabjolfaei2018fundamentals} such that $A_i=\{ j\in [5]:(j,i)\in \Gcal \}$. 
\item For any $i\in [5]$, $\Xc_i=\{0,1\}$ and $X_i\sim {\rm Bern}(p_i)$, where $p_i$ is uniformly randomly chosen from range $(0,1)$;
\item For any $i\in [5]$, $d_i=\dt_i\cdot H_{\infty}(\Xv_{G_i})$, where $\dt_i$ is uniformly randomly chosen from range $(0,1)$;
\item $P\subseteq [5]$ is randomly generated assuring that $|P|\le 2$. 
\end{itemize}
For each system, we compute the lower bounds $\Lc_{\rm max}^{\rm Thm. 1}$ and $\Lc_{\rm max}^{\rm Thm. 2}$ by Theorems \ref{thm:lower:bound:confusion} and \ref{thm:lower:bound:pm}, respectively. 
An interesting observation is that we have $\Lc_{\rm max}^{\rm Thm. 2}>\Lc_{\rm max}^{\rm Thm. 1}$ for only $2$ among the $500$ tested systems, while $\Lc_{\rm max}^{\rm Thm. 1}>\Lc_{\rm max}^{\rm Thm. 2}$ for all the $498$ remaining systems. 

We also compute the MaxL according to the privacy mechanism $P_{Y|\Xv_{[n]}}$ given by Algorithm \ref{alg:agglomerative}, denoted as $\Lc_{\rm max}^{\rm Alg. 1}$. 
Then we compute the ratio $R = \Lc_{\rm max}^{\rm Alg. 1}/\max( \Lc_{\rm max}^{\rm Thm. 1},\Lc_{\rm max}^{\rm Thm. 2} )$. 
A lower ratio $R$ (close to $1$) means that our converse and achievable results perform well and are quite close to the optimal $\Lc_{\rm max}^*$, while a higher ratio indicates bad performance.  
We summarize the values of $R$ from \Roycr{$500$} tests in Table \ref{tab:test}, from which we can see that the proposed techniques achieves a satisfactory level of performance for the majority of tested problems. \Roycr{Notably, we have $\Lc_{\rm max}^{\rm Thm. 1}=\Lc_{\rm max}^{\rm Alg. 1}$ and thus $R=1$ for $162$ among the $500$ tests. }

\newcommand{\opertypesmallone}[1]{\begin{minipage}{14mm}\centering\vspace{1mm} #1\vspace{1mm}\end{minipage}}
\newcommand{\opertypesmall}[1]{\begin{minipage}{9mm}\centering\vspace{1mm} #1\vspace{1mm}\end{minipage}}
\newcommand{\opertypesmalltwo}[1]{\begin{minipage}{15mm}\centering\vspace{1mm} #1\vspace{1mm}\end{minipage}}
\newcommand{\opertypemed}[1]{\begin{minipage}{70mm}\centering\vspace{1mm} #1\vspace{1mm}\end{minipage}}

\begin{table}[thb]
\begin{center}
\caption{Performance of the Lower Bounds versus Algorithm \ref{alg:agglomerative}.}
\begin{tabular}{|c|c|c|c|c|c|}
\hline
\opertypesmallone{Ratio, $R$}&\opertypesmall{$=1$}&\opertypesmall{$<1.05$}&\opertypesmall{$<1.1$}&\opertypesmall{$<1.2$}&\opertypesmall{$\ge 1.2$}\\
\hline
\opertypesmallone{Number of Systems} &\opertypesmall{$162$}&\opertypesmall{$401$}& \opertypesmall{$429$} &\opertypesmall{$460$} &\opertypesmall{$40$}\\
\hline
\end{tabular}\label{tab:test}
\end{center}
\end{table}

Future directions include improving the privacy mechanism and the converse results for the multi-terminal guessing problem \eqref{eq:model:tradeoff}, as well as studying the multi-terminal privacy-utility tradeoff using different privacy and utility measures.

\appendices

\section{Proof of Lemma \ref{lem:lower:bound:core}} \label{appendix:lem:core}

\Roycr{Note that the right hand side of \eqref{eq:model:ml:sibson} is equal to the Sibson mutual information of order $\infty$ \cite{sibson1969information}, $I_{\infty}^{\rm S}$, and hence we have
\begin{align*}
\Lc_{\rm max}(\Xv_Q\to Y|\Xv_P)&=I_{\infty}^{\rm S}(\Xv_Q;Y,\Xv_P)  \\
&\ge I(\Xv_Q;Y,\Xv_P)=I(\Xv_Q;Y|\Xv_P),
\end{align*}
where the inequality follows from \cite[Theorem 2]{verdu2015alpha}, and the last equality follows from source independence. }

It remains to show $\Lc_{\rm max}\ge \Roycr{\Delta_i}$ for any user $i\in [m]$ with $G_i\subseteq Q$.
For brevity, we drop the subscript $i$ remembering that $W,A,G$ stand for $W_i,A_i,G_i$, respectively.
Set
\begin{align}
\begin{array}{cc}
W_{P}=P\cap W, & W_{Q}=Q\cap W, \\
A_{P}=P\cap A, & A_{Q}=Q\cap A.
\end{array}
\nonumber   
\end{align}
Since $G\subseteq Q$, we have $P\cap G=\emptyset$, $Q\cap G=G$, and thus $P=W_P\cup A_P$, and $Q=W_Q\cup A_Q\cup G$.
We have
\begin{align}
&\Lc_{\rm max}(\Xv_Q\to Y|\Xv_P)   \nonumber   \\
&\ge \sum_{y,\xv_P} P_{Y,\Xv_P}(y,\xv_P)   \log \max_{\xv_Q} \frac{P_{Y,\Xv_P|\Xv_Q}(y,\xv_P|\xv_Q) }{P_{Y,\Xv_P}(y,\xv_P)}   \label{eq:lower:bound:core:proof:jensen} \\
&=\sum_{y,\xv_P}P_{Y,\Xv_P}(y,\xv_P)   \nonumber   \\
&\quad \log \big(   \max_{\xv_{W_Q},\xv_{A_Q}} (\frac{P_{Y,\Xv_P|\Xv_{W_Q},\Xv_{A_Q}}(y,\xv_P|\xv_{W_Q},\xv_{A_Q})}{P_{Y,\Xv_P}(y,\xv_P)} \nonumber  \\
&\quad  \cdot \max_{\xv_G} \frac{P_{Y,\Xv_P|\Xv_Q}(y,\xv_P|\xv_Q) }{P_{Y,\Xv_P|\Xv_{W_Q},\Xv_{A_Q}}(y,\xv_P|\xv_{W_Q},\xv_{A_Q})})   \big)  \nonumber   \\
&=\sum_{y,\xv_P}P_{Y,\Xv_P}(y,\xv_P)   \nonumber   \\
&\quad \log \big(   \max_{\xv_{W_Q},\xv_{A_Q}} (\frac{P_{Y,\Xv_P|\Xv_{W_Q},\Xv_{A_Q}}(y,\xv_P|\xv_{W_Q},\xv_{A_Q})}{P_{Y,\Xv_P}(y,\xv_P)} \nonumber  \\
&\quad  \cdot \max_{\xv_G} \frac{P_{Y,\Xv_A|\Xv_G}(y,\xv_A|\xv_G) }{P_{Y,\Xv_A}(y,\xv_A)})   \big)   \label{eq:lower:bound:core:proof:markov}   \\
&\ge \sum_{y,\xv_P,\xv_{A_Q}}P_{Y,\Xv_P,\Xv_{A_Q}}(y,\xv_P,\xv_{A_Q})   \nonumber   \\
&\quad \big(   \log \max_{\xv_{W_Q}} \frac{P_{Y,\Xv_P|\Xv_{W_Q},\Xv_{A_Q}}(y,\xv_P|\xv_{W_Q},\xv_{A_Q})}{P_{Y,\Xv_P}(y,\xv_P)} \nonumber  \\
&\quad  +\log \max_{\xv_G} \frac{P_{Y,\Xv_A|\Xv_G}(y,\xv_A|\xv_G) }{P_{Y,\Xv_A}(y,\xv_A)}   \big)   \label{eq:lower:bound:core:proof:max:ave:AQ}  \\
&=I(Y,\Xv_P;\Xv_{W_Q},\Xv_{A_Q})   \nonumber   \\
&\quad +\sum_{y,\xv_{A}}\hspace{-0.25mm}P_{Y,\Xv_{A}}(y,\xv_{A})\hspace{-0.25mm} \log \max_{\xv_G}\hspace{-0.25mm} \frac{P_{Y,\Xv_A|\Xv_G}(y,\xv_A|\xv_G) }{P_{Y,\Xv_A}(y,\xv_A)}   \label{eq:lower:bound:core:proof:max:ave:WQ}    \\
&\ge I(Y,\Xv_P;\Xv_{W_Q},\Xv_{A_Q})+D_i     \label{eq:lower:bound:core:proof:Dt:D}   \\
&=I(Y;\Xv_{A_Q}|\Xv_{P})+H(\Xv_{W_Q})+D_i=\Roycr{\Delta_i},    \label{eq:lower:bound:core:proof:last}
\end{align}
%
%
%
%
where
\begin{itemize}
\item \eqref{eq:lower:bound:core:proof:jensen} follows from Jensen's inequality;
\item \eqref{eq:lower:bound:core:proof:markov} follows from that for any $y\in \mathcal{Y}$, $\xv_{W\cup A}\in \Xc_{W\cup A}$, 
\begin{align*}
&\max_{\xv_G} \frac{P_{Y,\Xv_P|\Xv_Q}(y,\xv_P|\xv_Q) }{P_{Y,\Xv_P|\Xv_{W_Q},\Xv_{A_Q}}(y,\xv_P|\xv_{W_Q},\xv_{A_Q})}   \\
&=\max_{\xv_G} \frac{P_{Y,\Xv_{A\cup G}}(y,\xv_{A\cup G}) \hspace{-0.375mm}  \cdot \hspace{-0.375mm}   P_{\Xv_W|Y,\Xv_{A\cup G}}(\xv_W|y,\xv_{A\cup G}) }{P_{\Xv_G}(\xv_G) \hspace{-0.375mm}  \cdot  \hspace{-0.375mm}  P_{Y,\Xv_A}(y,\xv_A) \hspace{-0.375mm}  \cdot  \hspace{-0.375mm}  P_{\Xv_W|Y,\Xv_A}(\xv_W|y,\xv_A)}   \\
&\stackrel{(a)}{=} \max_{\xv_G} \frac{P_{Y,\Xv_A|\Xv_G}(y,\xv_A|\xv_G) }{P_{Y,\Xv_A}(y,\xv_A)},
\end{align*}
where (a) follows from the Markov chain $\Xv_W - (Y,\Xv_A) - \Xv_G$ as a result of the utility constraint \eqref{eq:model:utility:wi};
\item \eqref{eq:lower:bound:core:proof:max:ave:AQ} follows from replacing maximum over $\xv_{A_Q}$ with expectation over $P_{\Xv_{A_Q}|Y,\Xv_P}$, and Jensen's inequality;
\item \eqref{eq:lower:bound:core:proof:max:ave:WQ} follows from the fact that $\Xv_{W_Q}$ is a deterministic function of $(Y,\Xv_P,\Xv_{A_Q})$ according to \eqref{eq:model:utility:wi}; 
\item \eqref{eq:lower:bound:core:proof:Dt:D} follows from that 
\begin{align*}
&\sum_{y,\xv_{A}} P_{Y,\Xv_{A}}(y,\xv_{A}) \log \max_{\xv_G} \frac{P_{Y,\Xv_A|\Xv_G}(y,\xv_A|\xv_G) }{P_{Y,\Xv_A}(y,\xv_A)}   \nonumber   \\
&\ge \sum_{y,\xv_{A}} P_{Y,\Xv_{A}}(y,\xv_{A}) \log \frac{ \max\limits_{\xv_G}P_{Y,\Xv_A,\Xv_G}(y,\xv_A,\xv_G) }{\max\limits_{\xv_G}P_{\Xv_G}(\xv_G)\cdot P_{Y,\Xv_A}(y,\xv_A)}   \nonumber   \\
&={\mathbb E}_{P_{Y,\Xv_{A_i}}} \big[ \log r(\Xv_{G_i} \to Y|\Xv_{A_i}) \big]=D_i;
\end{align*}
\item \eqref{eq:lower:bound:core:proof:last} follows from source independence, as well as \eqref{eq:model:utility:wi}.
\end{itemize}

This concludes the proof.

\bibliographystyle{IEEEtran}
\bibliography{references}

\newcommand{\noopsort}[1]{}
\begin{thebibliography}{10}
\providecommand{\url}[1]{#1}
\csname url@samestyle\endcsname
\providecommand{\newblock}{\relax}
\providecommand{\bibinfo}[2]{#2}
\providecommand{\BIBentrySTDinterwordspacing}{\spaceskip=0pt\relax}
\providecommand{\BIBentryALTinterwordstretchfactor}{4}
\providecommand{\BIBentryALTinterwordspacing}{\spaceskip=\fontdimen2\font plus
\BIBentryALTinterwordstretchfactor\fontdimen3\font minus
  \fontdimen4\font\relax}
\providecommand{\BIBforeignlanguage}[2]{{%
\expandafter\ifx\csname l@#1\endcsname\relax
\typeout{** WARNING: IEEEtran.bst: No hyphenation pattern has been}%
\typeout{** loaded for the language `#1'. Using the pattern for}%
\typeout{** the default language instead.}%
\else
\language=\csname l@#1\endcsname
\fi
#2}}
\providecommand{\BIBdecl}{\relax}
\BIBdecl

\bibitem{Birk--Kol2006}
Y.~Birk and T.~Kol, ``Coding on demand by an informed source ({ISCOD}) for
  efficient broadcast of different supplemental data to caching clients,''
  \emph{{IEEE} Trans. Inf. Theory}, vol.~52, no.~6, pp. 2825--2830, Jun. 2006.

\bibitem{bar2011index}
Z.~Bar-Yossef, Y.~Birk, T.~Jayram, and T.~Kol, ``Index coding with side
  information,'' \emph{{IEEE} Trans. Inf. Theory}, vol.~57, pp. 1479--1494,
  2011.

\bibitem{arbabjolfaei2018fundamentals}
F.~Arbabjolfaei and Y.-H. Kim, ``Fundamentals of index coding,''
  \emph{Foundations and Trends{\textregistered} in Communications and
  Information Theory}, vol.~14, no. 3-4, pp. 163--346, 2018.

\bibitem{dau2012security}
S.~H. Dau, V.~Skachek, and Y.~M. Chee, ``On the security of index coding with
  side information,'' \emph{{IEEE} Trans. Inf. Theory}, vol.~58, no.~6, pp.
  3975--3988, 2012.

\bibitem{ong2016secure}
L.~Ong, B.~N. Vellambi, P.~L. Yeoh, J.~Kliewer, and J.~Yuan, ``Secure index
  coding: Existence and construction,'' in \emph{Proc. {IEEE} Int. Symp. on
  Information Theory ({ISIT})}, Barcelona, Spain, 2016, pp. 2834--2838.

\bibitem{mojahedian2017perfectly}
M.~M. Mojahedian, M.~R. Aref, and A.~Gohari, ``Perfectly secure index coding,''
  \emph{{IEEE} Trans. Inf. Theory}, vol.~63, no.~11, pp. 7382--7395, 2017.

\bibitem{liu:vellambi:kim:sadeghi:itw18}
\BIBentryALTinterwordspacing
Y.~Liu, Y.-H. Kim, B.~Vellambi, and P.~Sadeghi, ``On the capacity region for
  secure index coding,'' in \emph{Proc. {IEEE} Information Theory Workshop
  ({ITW})}, Guanzhou, China, Nov. 2018. [Online]. Available:
  \url{https://arxiv.org/abs/1809.03615}
\BIBentrySTDinterwordspacing

\bibitem{narayanan2018private}
V.~Narayanan, V.~M. Prabhakaran, J.~Ravi, V.~K. Mishra, B.~K. Dey, and
  N.~Karamchandani, ``Private index coding,'' in \emph{Proc. {IEEE} Int. Symp.
  on Information Theory ({ISIT})}, Vail, CO, 2018, pp. 596--600.

\bibitem{liu2020secure}
Y.~Liu, P.~Sadeghi, N.~Aboutorab, and A.~Sharififar, ``Secure index coding with
  security constraints on receivers,'' \emph{arXiv preprint arXiv:2001.07296},
  2020.

\bibitem{karmoose2019privacy}
M.~Karmoose, L.~Song, M.~Cardone, and C.~Fragouli, ``Privacy in index coding:
  k-limited-access schemes,'' \emph{{IEEE} Trans. Inf. Theory}, 2019.

\bibitem{issa2019operational}
I.~Issa, A.~B. Wagner, and S.~Kamath, ``An operational approach to information
  leakage,'' \emph{{IEEE} Trans. Inf. Theory}, 2019.

\bibitem{slonim2000agglomerative}
N.~Slonim and N.~Tishby, ``Agglomerative information bottleneck,'' in
  \emph{Advances in Neural Information Processing Systems (NIPS)}, 2000, pp.
  617--623.

\bibitem{makhdoumi2014information}
A.~Makhdoumi, S.~Salamatian, N.~Fawaz, and M.~M{\'e}dard, ``From the
  information bottleneck to the privacy funnel,'' in \emph{Proc. {IEEE}
  Information Theory Workshop ({ITW})}, Hobart, Australia, Nov. 2014, pp.
  501--505.

\bibitem{ding2019submodularity}
N.~Ding and P.~Sadeghi, ``A submodularity-based clustering algorithm for the
  information bottleneck and privacy funnel,'' in \emph{Proc. {IEEE}
  Information Theory Workshop ({ITW})}, Visby, Sweden, Aug. 2019, pp. 1--5.

\bibitem{renyi1961measures}
A.~R{\'e}nyi, ``On measures of information and entropy,'' in \emph{Proceedings
  of the 4th Berkeley symposium on mathematics, statistics and probability},
  vol.~1, no. 547, 1961.

\bibitem{alon2008broadcasting}
N.~Alon, E.~Lubetzky, U.~Stav, A.~Weinstein, and A.~Hassidim, ``Broadcasting
  with side information,'' in \emph{49th Annu. IEEE Symp. on Foundations of
  Computer Science (FOCS)}, Oct. 2008, pp. 823--832.

\bibitem{blasiak2011lexicographic}
A.~Blasiak, R.~Kleinberg, and E.~Lubetzky, ``Lexicographic products and the
  power of non-linear network coding,'' in \emph{52nd Annu. IEEE Symp. on
  Foundations of Computer Science (FOCS)}, Oct. 2011, pp. 609--618.

\bibitem{Arbabjolfaei--Bandemer--Kim--Sasoglu--Wang2013}
F.~Arbabjolfaei, B.~Bandemer, Y.-H. Kim, E.~Sasoglu, and L.~Wang, ``On the
  capacity region for index coding,'' in \emph{Proc. {IEEE} Int. Symp. on
  Information Theory ({ISIT})}, 2013, pp. 962--966.

\bibitem{yeung2008information}
R.~W. Yeung, \emph{Information theory and network coding}.\hskip 1em plus 0.5em
  minus 0.4em\relax Springer Science \& Business Media, 2008.

\bibitem{elgamal_yhk}
A.~{El Gamal} and Y.-H. Kim, \emph{Network Information Theory}.\hskip 1em plus
  0.5em minus 0.4em\relax Cambridge: Cambridge University Press, 2011.

\bibitem{sibson1969information}
R.~Sibson, ``Information radius,'' \emph{Zeitschrift f{\"u}r
  Wahrscheinlichkeitstheorie und verwandte Gebiete}, vol.~14, no.~2, pp.
  149--160, 1969.

\bibitem{verdu2015alpha}
S.~Verd{\'u}, ``$\alpha$-mutual information,'' in \emph{Proc. Inf. Theory Appl.
  Workshop (ITA)}, Feb. 2015, pp. 1--6.

\end{thebibliography}

\end{document}